\documentclass[11pt]{article} 
\usepackage{epsfig,amssymb,euscript} \usepackage{amsmath}
\newcommand{\ket}{\rangle} \newcommand{\bra}{\langle}
\newcommand{\eq}{\begin{equation}} \newcommand{\eeq}{\end{equation}}
\newcommand{\eqa}{\begin{eqnarray}} \newcommand{\eeqa}{\end{eqnarray}}

\newcommand{\Prob}{{\mathbf{P}}}

\newcommand{\be}{\begin{equation}} \newcommand{\ee}{\end{equation}}
\newcommand{\bea}{\begin{eqnarray}} \newcommand{\eea}{\end{eqnarray}}
\newcommand{\ba}{\begin{array}} \newcommand{\ea}{\end{array}}

  \def\bbox{{\,\lower0.9pt\vbox{\hrule \hbox{\vrule height 0.2 cm
\hskip 0.2 cm \vrule height 0.2 cm}\hrule}\,}} \newcommand{\dsl}{\pa
\kern-0.5em /}

\def\ds{\raise.15ex\hbox{/}\kern-.57em\partial}
\def\Ds{\,\raise.15ex\hbox{/}\mkern-13.5mu D}
\begin{document}

\begin{center}
\linespread{1.6}
\baselineskip=16pt {\Large\bf The Status of the Wave Function in Dynamical
Collapse Models}
\vskip 10.mm Fay Dowker$^{a}$ and Isabelle Herbauts$^{b}$\\
{\small \noindent $^a$ Blackett
Laboratory, Imperial College, London SW7 2AZ,
UK.\\
\noindent $^b$ Department of Physics, Queen Mary, University of
London, London E1 4NS, UK.\\ }
\end{center}
\hspace{8ex}


The idea that in dynamical wave function collapse models the wave
function is superfluous is investigated. Evidence is presented for
the conjecture that, in a model of a field theory on a 1+1
lightcone lattice, knowing the field configuration on the lattice
back to some time in the past, allows the wave function or quantum
state at the present moment to be calculated, to arbitrary
accuracy so long as enough of the past field configuration is
known.

\hspace{8ex}

\setcounter{equation}{0}

\section{INTRODUCTION}

The question of the status of the state vector, Psi, in standard 
text-book quantum mechanics has been a controversial issue 
since Bohr's and Einstein's time. 
Is the state vector a complete description of the  
physical state of a system, or is it incomplete and 
needing of completion with extra information, or 
does it represent a state of 
(someone or something's) knowledge of the system?   
The class of models known variously as ``dynamical collapse
models'' and ``spontaneous localisation models'' are observer
independent alternatives to standard quantum theory. 
We can ask: what happens to this question about the status of
the state vector? Is it immediately resolved by the 
formalism of the collapse models
 or is there still a question to answer?

The general structure of all dynamical collapse models is similar:
there is a state vector, Psi, which undergoes a stochastic
evolution in Hilbert space and there is a ``classical'' (c-number)
entity -- let's call it ``q-bar'' following Di\'osi
\cite{Diosi:1988a} -- with a stochastic evolution in spacetime.
The stochastic dynamics for the two entities -- Psi and q-bar --
are coupled together. The stochastic dynamics in Hilbert space
depends on which q-bar is realised in such a way as to tend to
drive Psi into an eigenstate of an operator (q-hat) that
corresponds to q-bar: this is the eponymous ``collapse'' in these
models. And the probability distribution for the realised values
of q-bar depends on Psi.

The choice of q-bar varies from model to model. In the original
GRW model \cite{Ghirardi:1986mt} and a proposed relativistic
version \cite{Tumulka:2004}, q-bar is a sequence of discrete
``collapse centres'' or spacetime events, in Di\'osi's model for
single particle quantum mechanics \cite{Diosi:1988a} q-bar is a
particle position (see, however, footnote 1), in Continuous
Spontaneous Localisation (CSL) models \cite{Pearle:1989,
Ghirardi:1990} q-bar is a scalar field. In all cases the c-number
entity q-bar is defined on spacetime and is therefore covariant in
essence.

The Bell ontology \cite{Bell:1987ii} for the GRW model states that
the collapse centres are the beables or real variables. The
analogous ontology for collapse models in general is that the
history of q-bar -- whatever it happens to be in the model -- is
real. Work by Di\'osi shows that any prediction about results of
macroscopic experiments and observations that can be made using
the expectation value of operator q-hat in state Psi, can also be
made, For All Practical Purposes (FAPP), using only knowledge
about q-bar, suitably regularised and coarse grained. Indeed, in
non-relativistic theories q-bar is equal to this expectation value
plus white noise with zero mean\footnote{This raises the objection
that the q-bar history is not really properly defined at all as it
contains a white noise term. One could fall back on the argument
that spacetime is widely expected to be fundamentally discrete and
this discreteness would provide a physical cutoff for the
frequency of the white noise. Or turn the argument around and say
that if the Bell ontology for collapse models is desirable, this
suggests the necessity of fundamental discreteness.}
 \cite{Diosi:1988a, Diosi:1988b}.
Put another way, suppose one has run one's computer simulations of
the collapse model up to a time well to the future of anything one
is interested in and in the computer memory is a history for Psi
traced out in Hilbert space and a (regularised) history for q-bar
traced out in spacetime. If the computer has a memory failure and
loses all information about Psi, the information about q-bar would
be enough, when suitably coarse grained, to make all the
macroscopic predictions that could be made from Psi.

An example is the lattice field theory \cite{Dowker:2002wm} that
is the subject of this paper. In \cite{Dowker:2004zn} it was
argued that a coarse graining of q-bar -- in this case q-bar is a
$\{0,1\}$-valued field on the lattice -- displays the same
structure, FAPP, as the coarse grained expectation value of the
field operator in the quantum state Psi.

In taking this point of view, that q-bar is real, we are forced to
address the question of the status of Psi. Di\'osi takes the view
that both Psi and q-bar are real \cite{Diosi:2004}. In this paper
we will investigate the possibility, raised explicitly by Kent
\cite{Kent:1989nk} for the GRW model, that Psi doesn't exist at
all -- that it is at most a convenience and conveys no information
that is not carried by the history of q-bar itself.

One can argue that there is already a way partially to demote the
quantum state in collapse models from its status as a really
existing thing to that of a ``dynamical law''. This view can be
taken in formalisms in which spacetime histories of the system are
primary (including for example consistent histories
\cite{Griffiths:1984rx, Omnes:1988ek, Gell-Mann:1989nu,
Hartle:1992as}, Sorkin's quantum measure theory
\cite{Sorkin:1994dt, Sorkin:1995nj} and Bohmian Mechanics, see
{\it e.g.} \cite{Durr:1995du}). If we consider a collapse model to
be a stochastic law for the q-bar histories then the quantum state
Psi can be formally relegated to the initial surface from which it
need never evolve. The initial state gives us the dynamical law
for the future q-bar histories in the form of the probability
distribution on them and is not itself real. However, we can, if
we know the q-bar history up to some spacelike surface, define an
``effective'' quantum state on that spacelike surface which tells
us how to calculate the probability distribution on q-bar events
to the future of the surface conditional on the known past
history. The ``evolution'' of this effective quantum state from
surface to surface (which is {\it precisely} the stochastic
process in Hilbert space mentioned above) is akin to a
``Bayesian'' updating -- on the actualisation of stochastic events
-- of the rule which gives the future probability distribution and
is not the evolution of something physical. On this view, the
quantum state is something we invent in order to render the
dynamics Markovian.

It would be desirable to go further than this. The initial state
on the initial surface hangs around like the smile of the Cheshire
Cat -- rather insubstantial but still persistently {\it there}.
Moreover, in the quest to make a relativistic collapse model, the
need to begin with a state defined on an initial surface breaks
Lorentz invariance. In this paper we elaborate on the conjecture
made in \cite{Kent:1989nk, Dowker:2002wm}, that in collapse models
even the initial state can be eliminated as a necessary part of
the theory (and the only information that remains from the state
is a classical distribution over superselection sectors). We
suggest that it can be replaced by an ``initial period of q-bar
history''. Knowing this initial period of history would allow the
calculation, FAPP, of an effective quantum state which could be
used to make predictions from then on.

In section 2 we briefly describe a collapse model for a field
theory on a 1+1 null lattice that we will use as a testing ground
for our conjecture. In this model, q-bar is a field configuration
of 0's and 1's on the lattice.  In section 3 we state the
conjecture and in section 4 we describe the simulations. The
results reported in section 5 suggest that if the field
configuration is known to a certain depth in time $T_{converge}$,
the state vector can be deduced FAPP from that configuration. Thus
the evolution of the field alone would be approximately Markovian
 on time scales larger than $T_{converge}$. Section 6
contains a summary and discussion.

\section{CAUSAL COLLAPSE MODEL ON A LIGHTCONE LATTICE}

We briefly review the spontaneous collapse model \cite{Dowker:2002wm}
that we will use to investigate the conjecture. We follow the
presentation of \cite{Dowker:2004zn} and refer to that paper for
further details.  The model is a modification of a
unitary QFT on a 1+1 null lattice, making it into a collapse
model by introducing local ``hits'' driving
the state into field eigenstates. The spacetime lattice is $N$
vertices wide and periodic in space, extends to the infinite future,
and the links between the lattice points are left or right going null
rays. A spacelike surface $\sigma$ is specified by a sequence of
$N$ leftgoing links and $N$ rightgoing links cut by the surface;
examples of spatial surfaces are shown in figure 1.
We assume an initial spacelike surface $\sigma_0$.
\begin{figure}[p]
\epsfxsize=10cm \centerline{\epsfbox{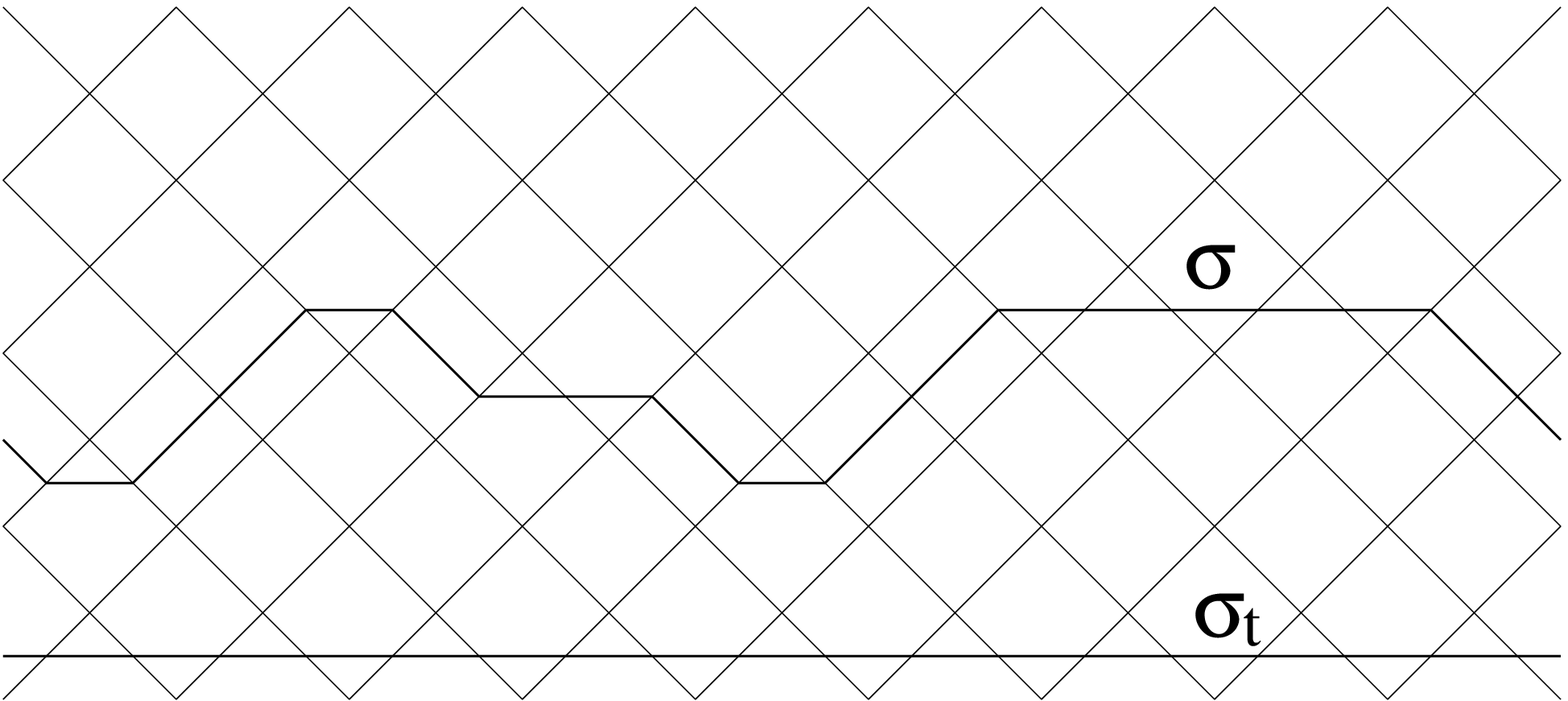}}
\caption{The light cone lattice. $\sigma_t$ is a constant time
surface and $\sigma$ is a generic spacelike surface.}
\label{fig:lattice}
\end{figure}

An assignment
of labels to the vertices to the future of $\sigma_0$, $v_1, v_2, \dots$,
 is called ``natural'' if $i< j$ whenever $v_i$ is to the causal
past of $v_j$. A natural labelling is equivalent to a linear
extension of the (partial) causal order of the vertices. A natural
labelling, $v_1, v_2, \dots$ is also equivalent to a sequence of
spatial  surfaces, $\sigma_1, \sigma_2, \dots$ where the surface
$\sigma_k$ is defined such that between it and $\sigma_0$, lie
exactly the vertices $v_1, \dots v_k$. One can think of the
natural labelling as giving an ``evolution'' rule for the
spacelike surfaces: as each vertex event $v_k$ occurs, the surface
creeps forward by one ``elementary motion'' across that vertex.
For any natural labelling and any $k$, the finite set of vertices
$\{v_1, v_2, \dots v_k\}$ is a {\it stem}, a finite set that
contains its own causal past.

The local field variables $\alpha$ live on the links. These field
variables take only two values $\{0,1\}$, so that on each link
there is a qubit Hilbert space spanned by these two states. We
denote by $\{\alpha_{R_k},\alpha_{L_k}\}$ ($\alpha_{v_k}$ for
short) the values of the field variables on the two outgoing links
(to the right $R$ and to the left $L$) from vertex $v_k$. (Note,
in paper 1 we used the hatted symbol $\hat\alpha$ to denote the
actual value of the field variable but here we will use the
unhatted $\alpha$.) One can, colloquially, consider the field
values $0$ and $1$ to represent the absence or presence (resp.) of
``bare particles'' on the lattice.

A quantum state $|\psi_n \rangle$ on surface $\sigma_n$ is
an element of the $2^{2N}$ dimensional Hilbert space $H_{\sigma_n}$
which is a tensor
product of the 2N 2-dimensional
Hilbert spaces on each link cut by $\sigma_n$.
The
basis vectors (the ``preferred basis'') of this Hilbert space are
labelled by the possible field configurations on $\sigma_n$, namely
the $2N$-element bit strings $\{0,1\}^{2N}$. We will often
refer to the number of 1's in the bit string labelling an eigenstate
as the number of particles in that state. The Hilbert space is
the direct sum of $2N + 1$ sectors each of fixed particle number.
We identify  the Hilbert spaces on different surfaces in the obvious way
using the field basis. At each vertex $v_k$,
there is a local evolution law which is given by a 4-dimensional
unitary ``R-matrix'' $U(v_k)$ (4-dimensional
because it evolves from the two ingoing links
the two outgoing links).
For this paper we choose these R-matrices to be uniform across the lattice.
(One can simulate external interventions by fiddling with
the R-matrices.)

In the standard text-book unitary theory, one postulates the existence of
an external measuring agent and then this formalism
can be used to predict the results of sequences of measurements
of the field. One way to do this is to
identify projectors:
$P(\alpha_{v_k})$ projects onto the subspace of the Hilbert space
spanned by the basis vectors in which the field values at vertex $v_k$
are $\alpha_{v_k}$ (recall there are two links outgoing from $v_k$ and
so $\alpha_{v_k}$ is really two values).

Then the joint probability
distribution for the agent
to  \textit{measure} a particular field configuration
$\{\alpha_{v_1}, \alpha_{v_2}, \dots \alpha_{v_n}\}$
on the lattice between the hypersurfaces $\sigma_0$ and
$\sigma_n$ is:

\eq \Prob^{\mbox{standard QM}} (\alpha_{v_1}, \alpha_{v_2}, \dots
\alpha_{v_n})= || P(\alpha_{v_n})U(v_n) \dots
P(\alpha_{v_1})U(v_1) |\psi_0\rangle||^2 \, . \label{eq:prob_stan}
\eeq

This probability rule is independent of the linear ordering
of the vertices and depends only on their causal order. The rule
evades the potential danger of violating relativistic causality
described in \cite{Sorkin:1993gg} in two ways: the causal structure
of the vertices of the lattice is a partial order from the start (no
transitive completion is required) and the field variables being
measured are completely local quantities.

Inspired by the GRW model with Bell's ontology, this unitary
quantum field theory requiring external agents can be turned into
an observer independent theory for a closed system by replacing
the projection operators for measurements in \eqref{eq:prob_stan}
by positive operators (for ``unsharp measurements'') and adopting
the resulting formula as the probability that the corresponding
field configuration {\it occurs}. More precisely, we define on
each link ({\i.e.} on each 2-dimensional Hilbert space associated
with a link) the two operators $J_0$ and $J_1$ where
\be J_0 = \frac{1}{\sqrt{1+ X^2}}\begin{pmatrix} 1&0\\ 0&X\\
\end{pmatrix}\;, \quad J_1 = \frac{1}{\sqrt{1+ X^2}}\begin{pmatrix}
X&0\\ 0&1\\
\end{pmatrix}\;
\label{eq:jumps}
\ee
with $0\le X\le 1$. Note that $J_0^2 + J_1^2 = 1$ and this is a
positive operator valued measure. Then we define the jump operator
$J({\alpha}_{v_k})$ on the 4-dimensional Hilbert space on the
outgoing links from $v_k$ as the tensor product of the two
relevant 2-dimensional jump operators, {\it e.g.} when
${\alpha}_{v_k} = (0,0)$, $J({\alpha}_{v_k})= J_0 \otimes J_0$. We
promote $J({\alpha}_{v_k})$ to an operator on the Hilbert space of
any spatial surface containing those two links by taking the
tensor product with the identity operators on all the other
components of the full Hilbert space. The probability of the field
configuration $\{{\alpha}_{v_1}, \dots {\alpha}_{v_n}\}$ is given
by
\be\label{eq:prob_coll} \Prob({\alpha}_{v_1}, \dots
{\alpha}_{v_n}) = \Vert J({\alpha}_{v_n})U(v_n)\dots
J({\alpha}_{v_1})U(v_1) |\psi_0 \rangle \Vert^2\; .  \ee
From this we can see the importance of the fact the jump
operators form a positive operator valued measure, which 
ensures consistency:
\begin{align}\label{eq:consistent} \Prob({\alpha}_{v_1}, \dots
{\alpha}_{v_{n-1}}) =& \sum_{\alpha_{v_n}} \Prob({\alpha}_{v_1}, \dots
{\alpha}_{v_n})\\
=&  \sum_{\alpha_{v_n}} \langle \psi_0 \vert \dots J(\alpha_{v_n}) 
J(\alpha_{v_n}) \dots \vert \psi_0 \rangle\\
=& \Vert J({\alpha}_{v_{n-1}})U(v_{n-1})\dots
J({\alpha}_{v_1})U(v_1) |\psi_0 \rangle \Vert^2\; .
\end{align}

Again, \eqref{eq:prob_coll} depends only on the (partial) causal order of the
vertices because any other choice of natural labelling of the same
vertices gives the same result. These probabilities of the field
configurations on all stems are enough, via the standard methods
of measure theory, to define a unique probability measure on the
sample space of all field configurations on the semi-infinite
lattice.

We stress that whereas equation \eqref{eq:prob_stan} is the
probability for \textit{measuring} a particular field
configuration in standard unitary quantum theory, equation
\eqref{eq:prob_coll} is interpreted, in the Bell ontology, as the
probability for the field to \textit{be} in that configuration.
The full content of the theory is the probability distribution
\eqref{eq:prob_coll} on possible field configurations, dependent
on an initial state $|\psi_0\rangle$.

The {\it state} on the hypersurface $\sigma_n$ that is reached after
the elementary motions over vertices $v_1, \dots v_n$ and the field
values $\{{\alpha}_{v_1}, \dots {\alpha}_{v_n}\}$ have been
realised is the normalised state
\be\label{eq:state} |\psi_n \rangle= \frac{J({\alpha}_{v_n})U(v_n)\dots
  J({\alpha}_{v_1}) U(v_1)|\psi_0\rangle} {\Vert
  J({\alpha}_{v_n})U(v_n)\dots J({\alpha}_{v_1})
  U(v_1)|\psi_0\rangle \Vert} \,.\ee
Thus, the probability for state \eqref{eq:state} on hypersurface 
$\sigma_n$ is \eqref{eq:prob_coll}.
In order to make predictions about the field on the lattice
to the future of $\sigma_n$ -- conditional on the past values
$\{{\alpha}_{v_1}, \dots {\alpha}_{v_n}\}$ -- it is sufficient
to know $|\psi_n\rangle$. Indeed, the conditional probability of
$\{{\alpha}_{v_{n+1}}, \dots {\alpha}_{v_{n+m}}\}$
is given by
\be\label{eq:prob_cond} \Prob({\alpha}_{v_{n+1}}, \dots
{\alpha}_{v_{n+m}}) = \Vert J({\alpha}_{v_{n+m}})U(v_{n+m}\dots
J({\alpha}_{v_{n+1}}) U(v_{n+1})|\psi_n\rangle \Vert^2\; .  \ee
The state vector provides these conditional probabilities, and is
therefore a convenient way of keeping the probability distribution up to date,
given past events.

\section{THE STATUS OF THE WAVE FUNCTION: ``AN EXECUTIVE SUMMARY''?}

We are interested in investigating the possibility of doing away with the
quantum state entirely as a fundamental concept in collapse models
and will be using the lattice model described above as a test case.
As mentioned in the introduction, we can relegate the quantum state
to a state, $|\psi_0\rangle$, on an
initial surface $\sigma_0$ from where it acts as a ``dynamical
law'', specifying the probability distribution on field configurations
to the future of $\sigma_0$. Can we weaken even this status?

We make the conjecture that, if the field configuration
is known between $\sigma_0$ and $\sigma_n$, then even if the
state on $\sigma_0$ is not known, the state on $\sigma_n$ is
calculable up to a correction that goes to zero as $n\rightarrow \infty$.
This would mean that although the evolved state on
$\sigma_n$ {\it a priori} depends on both the initial state
on $\sigma_0$ and on the field values that have actually occurred
in between, its dependence on $|\psi_0\rangle$ dies away as time
goes on until all we need to know
to make predictions, FAPP, is the field configuration
back to a certain depth in time.
 We would then have an interpretation not only assigning reality to a field
 configuration in spacetime but further demoting the wave function
 by denying it a role as a \textit{necessary} entity to the theory: the
 state $|\psi_n\rangle$ can be deduced FAPP
from the field configuration to the past of $\sigma_n$ to some
depth in time (or exactly if the whole infinite past history
is known) and becomes an ``executive
 summary'' of the past reality containing no independent
 information.

More precisely, let $|\psi^1_0\rangle$ and $|\psi^2_0\rangle$ be
two states on $\sigma_0$. Then they give, according to
\eqref{eq:prob_coll}, two probability distributions, $\Prob_{1}$
and $\Prob_{2}$ on field configurations to the future of
$\sigma_0$. Choose any linear ordering of the vertices to the
future of $\sigma_0$, $v_1, v_2, \dots$. Adopt the notation
$\alpha(n)$ for a field configuration between $\sigma_0$ and
$\sigma_n$, and $|\psi^a_n, \alpha(n) \rangle$ for the state on
$\sigma_n$ that arises from $|\psi^a_0\rangle$ on $\sigma_0$ after
$\alpha(n)$ has happened ($a = 1,2$).

Conjecture: There exists a complex phase $\lambda$ such that
\eq
\Vert |\psi^2_n, \alpha(n) \rangle - \lambda |\psi^1_n, \alpha(n)
\rangle \Vert \rightarrow 0 \hspace{5pt}\mbox{as}\hspace{5pt} n\rightarrow \infty
\eeq
 for all
$\alpha(n)$ except those which almost surely do not occur
according to both $\Prob_{1}$ and $\Prob_{2}$.


Conjecture (density matrix form):

\eqa
\lefteqn{\Vert \sum_{\alpha(n)} \Prob_1(\alpha(n)) |\psi^1_n,
\alpha(n) \rangle \langle \psi^1_n, \alpha(n)|} \nonumber \\
  & &- \sum_{\alpha(n)}\Prob_1(\alpha(n)) |\psi^2_n, \alpha(n) \rangle \langle \psi^2_n,\alpha(n)|\Vert \rightarrow 0 \hspace{5pt}\mbox{as}\hspace{5pt} n\rightarrow \infty
\eeqa

 where
$\Vert \cdot\Vert$ is the operator norm, and similarly with 1 and
2 interchanged.

Note that we already know that the conjectures cannot be true
strictly as stated because of the possible existence of
``superselection sectors'' in the Hilbert space. For example, the
jump operators $J$ preserve particle number and if the R-matrices
do so also (this is the case we will study in detail in the next
section) then a state in the $k$-particle sector can never
approach a state in the $l$-particle sector if $k \ne l$. If the
R-matrices preserved only particle number mod-2 (by allowing pair
creation and annihilation of particles) then there would be two
superselection sectors (even and odd particle number). It should
be noted that even if there is a conserved quantity -- particle
number, say -- this quantity is conserved in the state vector but
not in the realised field configuration. We expect however that
the ``conservation law'' will be reflected in the probability
measure in the sense that a suitable property of the coarse
grained field configuration will be predicted with probability
close to one.

When there are superselection sectors, an initial quantum state
corresponds to a classical probability distribution over the
sectors and a quantum state in each sector, in the familiar way.
Without loss of generality therefore we will assume in what
follows that we are restricted to a single superselection sector
and the conjectures apply to each superselection sector
individually becoming, effectively: two states in the same
superselection sector tend to each other up to a phase for all
histories except for a set of histories which has measure zero in
the probability measure of both states.

\section{THE SIMULATIONS}

We sought evidence for the conjecture in the following way.
We chose a unitary R-matrix, uniform across the lattice,
of the following form:
\be{R = \bordermatrix{ &
 \scriptscriptstyle{\phantom{\nearrow\nwarrow}} &
 \scriptscriptstyle{\phantom{\nearrow}\nwarrow} &
 \scriptscriptstyle{\nearrow\phantom{\nwarrow}} &
 \scriptscriptstyle{\nearrow\nwarrow} \cr
 \scriptscriptstyle{\phantom{\nwarrow\nearrow}} & 1 & 0 & 0 & 0 \cr
 \scriptscriptstyle{\phantom{\nwarrow}\nearrow} & 0 & i\sin\theta &
 \cos\theta & 0 \cr \scriptscriptstyle{\nwarrow\phantom{\nearrow}} & 0
 & \cos\theta & i\sin\theta & 0 \cr
 \scriptscriptstyle{\nwarrow\nearrow} & 0 & 0 & 0 & 1 \cr } \; .
\label{eq:rmat}
} \ee
This gives a particle number preserving dynamics for the
state, since the hit operators $J$ also preserve particle
number.

We chose $\sigma_0$ to be a constant time surface and
we chose two initial states, $|\psi^1_0\ket$ and $|\psi^2_0\ket$ (in the same
superselection sector, which here meant the same particle number sector).
We generated, at random according to the probability distribution
$\Prob_1$ or $\Prob_2$
field configurations to the future of $\sigma_0$. For each of these
field configurations, $\alpha(M)$ (where $M$ was large enough
for the calculation in hand) we
calculated the two states $|\psi^a_n, \alpha(n)
\ket$, $a=1,2$ on the surface $\sigma_n$ which is the n${}^{th}$ surface in
a sequence of surfaces chosen according to the stochastic rule ``choose the
next elementary motion at random with uniform probability from those
possible.'' This is not a covariant rule -- it is equivalent to a
probability distribution on linear extensions of the partial
order on the whole future lattice but it does not
give each equal weight --
and moreover a covariant, Markovian rule does exist \cite{Brightwell:2004}
but we made the choice for ease of calculation. We will comment on
what significance this has for our results below.

We would like to show that the two states  $|\psi^a_n, \alpha(n)
\ket$, $a=1,2$ become close, up to a phase, as $n$ gets large and
more precisely we would
like to know how the difference behaves with $n$. One can argue that
it is not the states themselves that should be compared but the
probability distributions for the field variables that they
produce. Indeed, in an interpretation in which only the field is
real, it is only this probability distribution and not the state itself
which has physical import.

In principle the entities that should be compared are the two
probability distributions, for states 1 and 2, over field
configurations to the future of any surface $\sigma$, given the
values of $\alpha( M)$ lying to the past of $\sigma$. This is
calculationally impractical and we used two simplifying
strategies. First, we sampled the space of all surfaces by
choosing a sequence $\sigma_1, \sigma_2, \dots$ according to the
rule described above. This rule does not sample uniformly in the
space of surfaces, as mentioned above, and an improvement of our
scheme would be to determine and then implement the covariant rule
which does. Second, we compared the probabilities, not for the
whole future field configuration but only for the value 1 on each
of the two outgoing links from $v_n$ for each $n$.

An important point is that, as discussed in \cite{Dowker:2004zn},
the interesting physical regime for these models is when the
parameter $X$ is close to one, alternatively when $\epsilon \equiv
1-X$ is close to zero. This means that the hits are very gentle
and superpositions of microscopically different states will last
for a long time. In this case, however, the conditional
probability of a $1$ on each link becomes very close to $1/2$,
indeed it is equal to $1/2 + O(\epsilon)$. (Here we clearly see
the white noise term in the field configuration that is to be
expected from Diosi's work.) So for small epsilon the
probabilities will be close, whether or not the states are coming
close to each other. Indeed, let the link in question be denoted
$l$ and suppose, at some stage in the dynamics, $l$ is one of the
outgoing links from the vertex that has just been evolved over.
Let the state on the current spacelike surface through $l$ be
denoted schematically by
\be \label{current.eq}
\vert \Psi \rangle = a \vert 0 \rangle + b \vert 1 \rangle \ee
where $\vert 0 \rangle$ ($\vert 1 \rangle$) is short hand for the
normalised superposition of all the terms in the state in which the
value of the field on $l$ is 0 (1).  The probability that the field
will be 1 on $l$ (conditional on the past evolution to that stage) is
\be \label{probone.eq} \frac{|a|^2 X^2 + |b|^2}{1 + X^2}=  \frac{X^2}{1 + X^2} - |b|^2 \frac{1-X^2}
{1+X^2}\; . \ee
So, for the difference between the probability for a 1
on link $l$ in state 1 and in state 2 we will obtain:
\be
(|b_1|^2 - |b_2|^2) \frac{1-X^2}
{1+X^2}\; .
\ee
When $\epsilon$ is small this becomes
\be
(|b_1|^2 - |b_2|^2) (\epsilon + {\cal{O}}(\epsilon^2))
\; .
\ee

From this we see that the appropriate quantity to calculate for
each link is $|b_1|^2 - |b_2|^2$: that gives a measure of
the difference of the probability distributions that
affects the coarse grained, renormalised field
configuration (see \cite{Dowker:2004zn})
and indeed it is a measure of the difference between the
states themselves.

Thus, we calculated for each vertex $v_n$ (recall the sequence
$v_1, v_2, \dots$ is equivalent to a linear extension and is the
one chosen at random by our evolution rule described above) and
for each outgoing link, $l$, from $v_n$, the quantity $ |b_1|^2 -
|b_2|^2$ which we denote by $B(l)$. Of course this is a very
approximate
 measure of the difference between the two states
and to overcome this, we took the sum of this quantity over every
link in a block, a certain number, $m$, of lattice time steps long
and the width of the whole spatial lattice:
\be\label{Bt.eq}
 B_{m}(t) \equiv \sum_{ l} B(l)
\ee
where the sum is over all links
with lattice time coordinate from $t$ through
$t+m -1$ (the lattice time
step is 1).
 Our convergence
criterion was $B_{m}(t) < \delta$ and we define the convergence
time $T_{c}$ to be the smallest time such that $B_{m}(t)\le
\delta$, $\forall t> T_{c}$.

With the help of numerical simulations on 8, 9 and 10 vertex
lattices we studied the dependence of $T_{c}$ on $\epsilon$, on
particle number, on $\theta$ and on different types of initial
state within fixed particle number sectors. In total about  600
simulations were run. We also studied the convergence of states
for field configurations not generated according to the
probability distributions from either state, for example the field
configuration  (a) of all 1's, (b) of all 0's and (c) randomly
generated with uniform probability distribution of 1/2 for a 1 on
each link. We failed to find convergence only in the cases (a) and
(b) mentioned above when the field configuration was all 1's or
all 0's which is consistent with the conjecture because they
almost surely do not occur in $\Prob_1$ and in $\Prob_2$.
Convergence occurred but was slower for the field configurations
of type (c) than for those generated by (and therefore likely in)
the probability distibutions of states 1 or 2.

In our simulations we were limited as to lattice size by the
exponential growth of the problem in vertex number, and it is at
present unclear whether the limited size of the lattice has important
implications for our results, in particular the question whether the
periodic boundary conditions of the lattice stimulate convergence
remains open.

\section{Results}

\begin{figure}[p]
\epsfxsize=14cm \centerline{\epsfbox{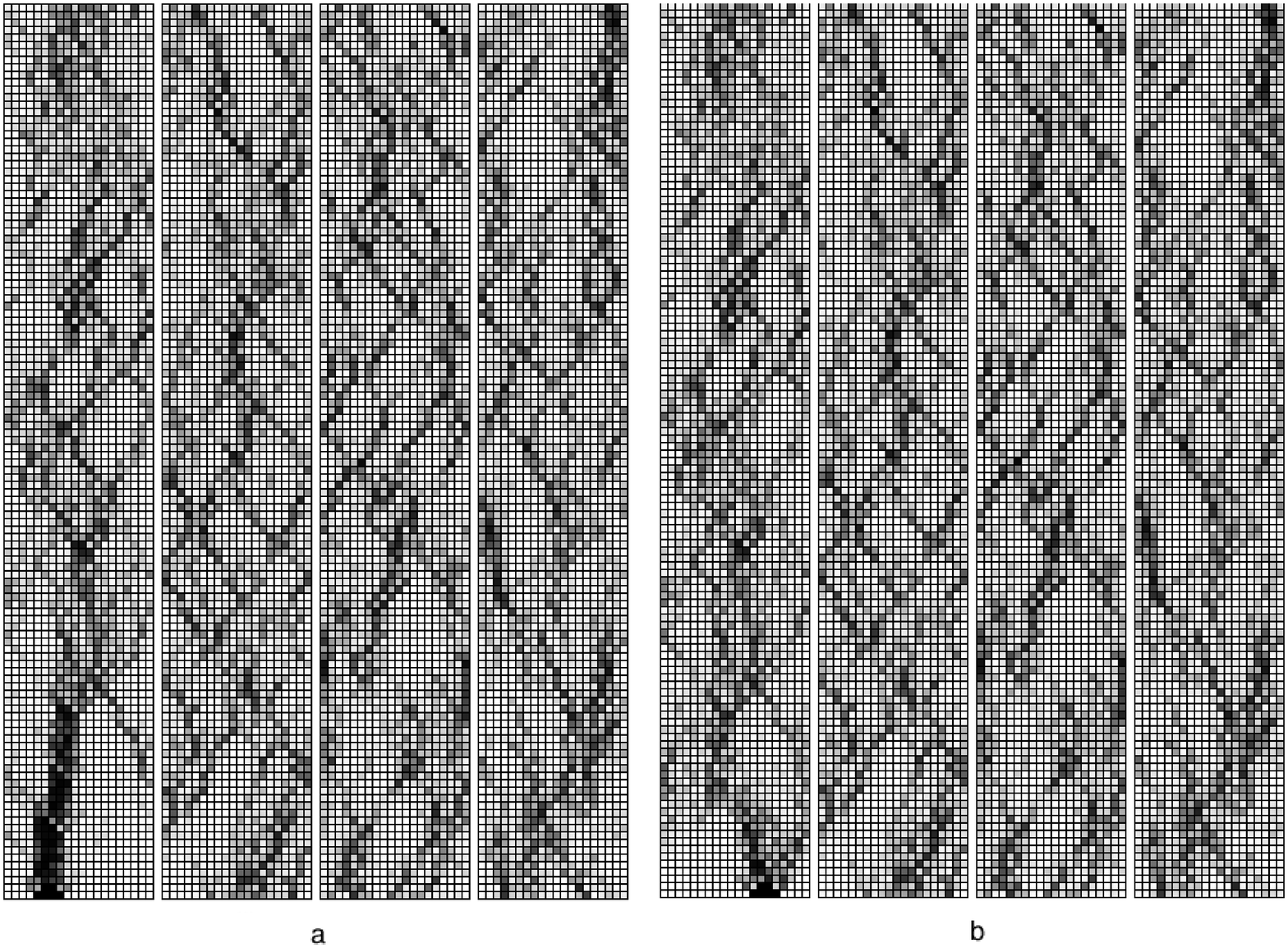}}
\caption{(a) and (b) are plots of the evolution of four-particle
eigenstates $|\psi_1\ket$ and $|\psi_2\ket$ respectively, in a
field configuration generated according to the probability
distribution of $|\psi_1\ket$, with $X=0.65$ and $\theta=
0.26\pi$.} \label{fig:states}
\end{figure}

To begin by giving a flavour of the kind of simulations run, the
convergence of two initial states is illustrated in
 figure \ref{fig:states}. Each cell corresponds to a
single link of the lattice (so there are twice as many cells
across the lattice width as vertices) and the darkness of the cell
is (positively) proportional to $|b|^2$ (see equation
\eqref{current.eq}). In plot (a) we show the evolution of state 1,
which begins as an eigenstate with 4 particles on the left hand
side of the lattice in the leftmost panel, time proceeds up the
page and then the lattice continues at the bottom of the next
panel and so on. Plot (b) shows the evolution of state 2, which
begins as a state with 4 particles on the right hand side of the
lattice. The parameters for the evolution are $X=0.65$ and
$\theta=0.26\pi$. Comparing (a) and (b) it can be seen that the
plots are indistinguishable from halfway up the first panel in
each. The plots after this time are somewhat superfluous but we
show them to emphasize that we checked that the convergence
persists long after our convergence criterion is reached.

\begin{figure}[p]
\epsfxsize=12cm \centerline{\epsfbox{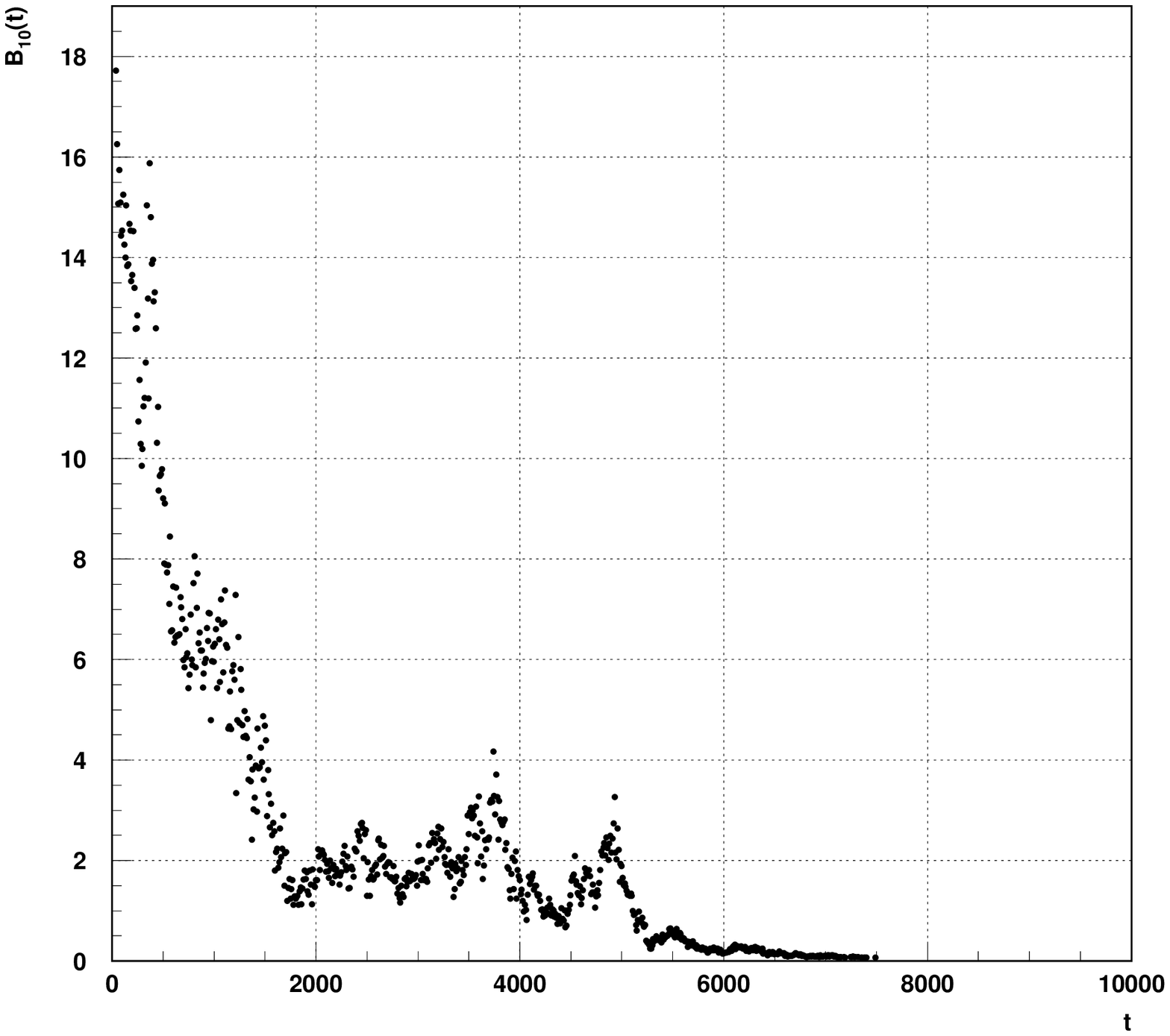}}
\caption{Plot of $B_{10}(t)$, against lattice time $t$, for
$X=0.95$, $\theta= 0.1\pi$ and two different one-particle
eigenstates for initial states.} \label{B10.fig}
\end{figure}

Figure \ref{B10.fig} shows the quantity $B_{10}(t)$ defined in
equation \eqref{Bt.eq} plotted against lattice time $t$, for an 8
vertex run with $X=0.95$, $\theta= 0.1\pi$ and initial states
which are two different one-particle eigenstates. It shows a
pleasingly sharp falloff to zero.

\begin{figure}[p]
\epsfxsize=16cm \centerline{\epsfbox{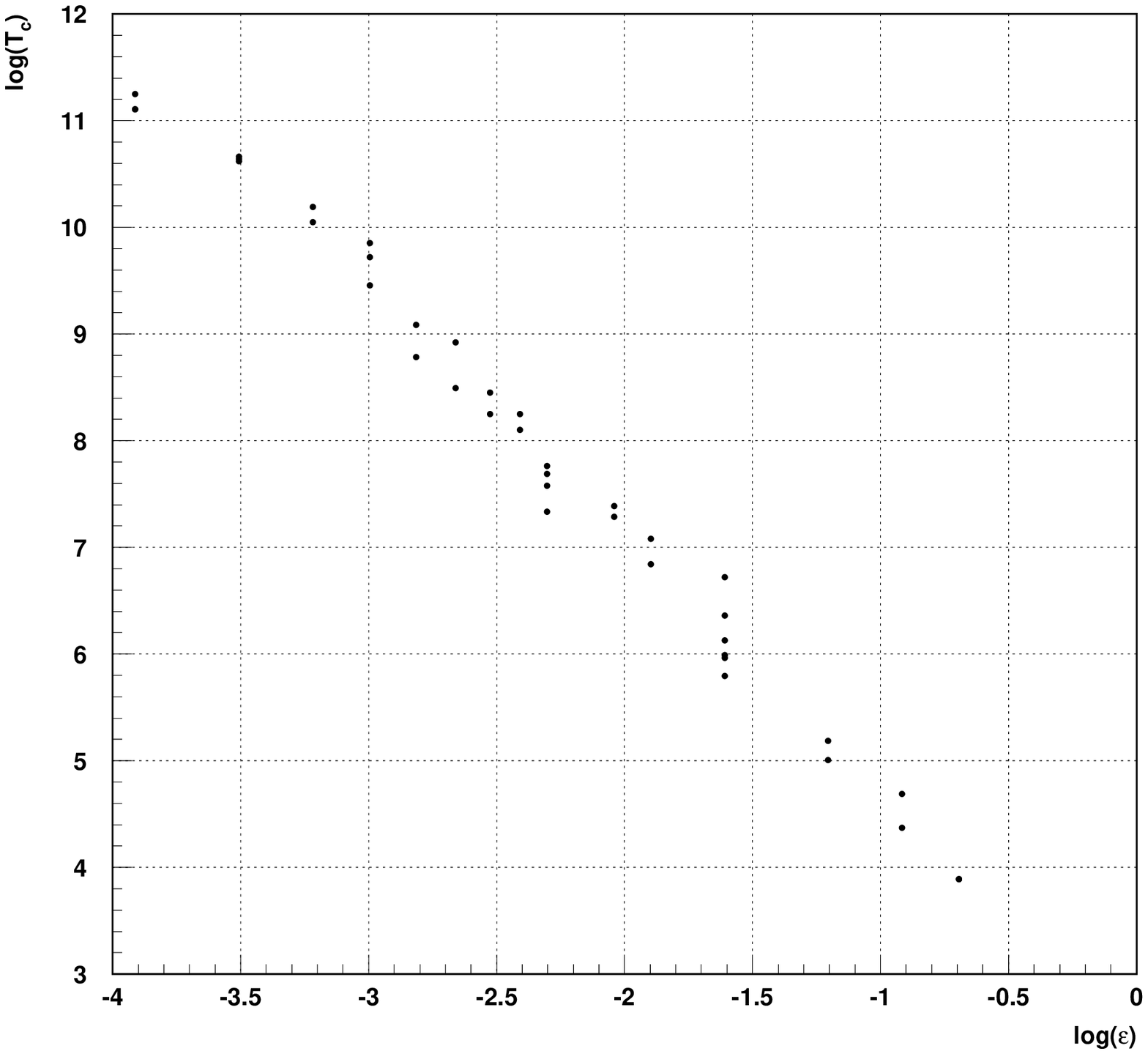}} \caption{Plot
of $\log(T_{c})$ against $\log(\epsilon)$} \label{plot1.fig}
\end{figure}

Figure \ref{plot1.fig} is a plot of $\log(T_{c})$ against
$\log(\epsilon)$ for many 8 vertex runs of varying $\epsilon$,
where we chose $B_{10}$ as our measure of difference and $\delta =
10^{-4}$ to define the convergence time. The other parameters were
$\theta = 0.1\pi$ and $|\psi_1\ket$ and $|\psi_2\ket$ are two
fixed one-particle eigenstates. The field configuration is chosen
according to the probability distribution from $|\psi_1\ket$. The
plot is consistent with a dependence of \eq T_{c}\propto
\frac{1}{\epsilon^2} \eeq

As a consequence of this, in a continuum limit in
which the lattice spacing $a \rightarrow 0$ and
$\epsilon = {\cal{O}}(\sqrt{a})$, the ``physical''
convergence time, $aT_{c}$ would tend
to some finite non-zero value.

\begin{figure}[p]
\epsfxsize=16cm \centerline{\epsfbox{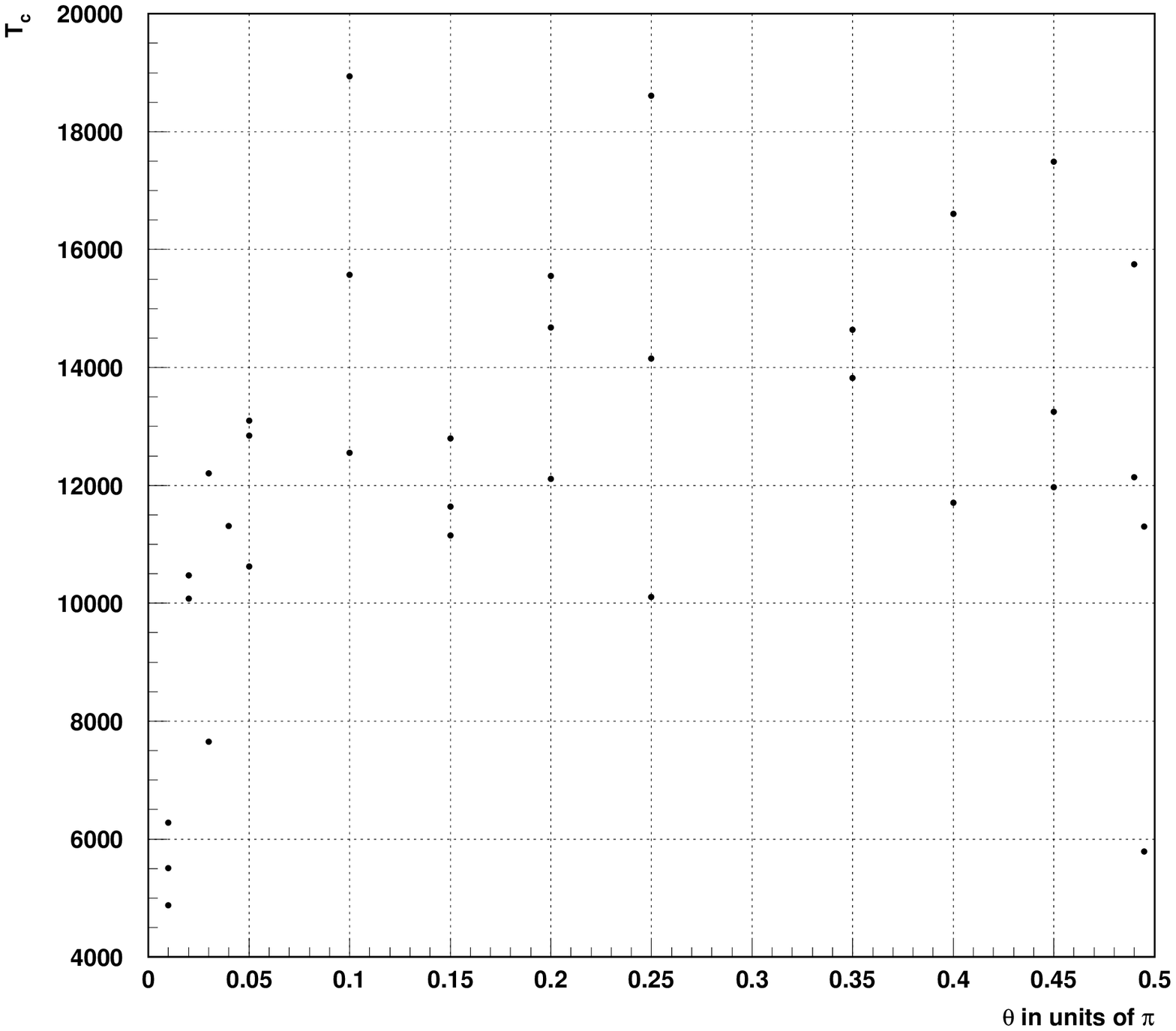}}
\caption{Convergence time vs $\theta$.} \label{plot2:fig}
\end{figure}

A plot of the convergence time (defined by $B_{10}$ and $\delta =
10^{-4}$) against $\theta$ is given in figure \ref{plot2:fig} for
$X = 0.95$ and fixed initial one-particle eigenstates.

The plot is difficult to interpret. It seems particularly odd when one
realises that for $\theta = \pi/2$ and $\theta = 0$ (the two
limits of the range of $\theta$ shown) the R-matrix is such
that it does not introduce any superpositions into the states.
Indeed the evolution is completely deterministic: for
$\theta=\pi/2$ an initial one-particle field eigenstate
remains essentially constant -- just acquiring a phase of $i$
at each lattice time step -- and for $\theta = 0$ it
propagates at the speed of light along the null direction  it
starts off in.
The evolution in these cases, therefore, does not ``mix'' the Hilbert space
and states 1 and 2 can never converge.

The plot, however, suggests that for values of $\theta$ close to
these limiting ones, convergence is faster than for $\theta$ in
the middle of the range, and indeed the convergence time is
tending to zero. We speculate that this has something to do with
the competing effects of the mixing by the R-matrices and the
converging effect of the hits. If we imagine starting with two
states which have support over the whole of the one-particle
sector of Hilbert space, the harder the hits, the faster the
states will converge. In the extreme case, if $\epsilon = 1$ then
the hit operators are projectors, state 1 will collapse into one
of the eigenstates after one time step, the field configuration
will be the one given by that eigenstate and state 2 will be
forced into that state also. When $\epsilon$ and $\theta$ vary,
there is a competition between the driving towards eigenstates by
the hits and the mixing (introduction of superpositions) by the
R-matrices. In the runs plotted here, we kept $\epsilon$ fixed so
the strength of the hits does not vary but as $\theta$ tends
towards the two limiting values it could be that the R-matrix
evolution loses the competition. If the hits drive state 1 very
quickly into an eigenstate, then as long as there's been enough
mixing so that there is even a tiny amplitude for that eigenstate
in state 2, there will be convergence.

\begin{figure}[p]
\epsfxsize=16cm \centerline{\epsfbox{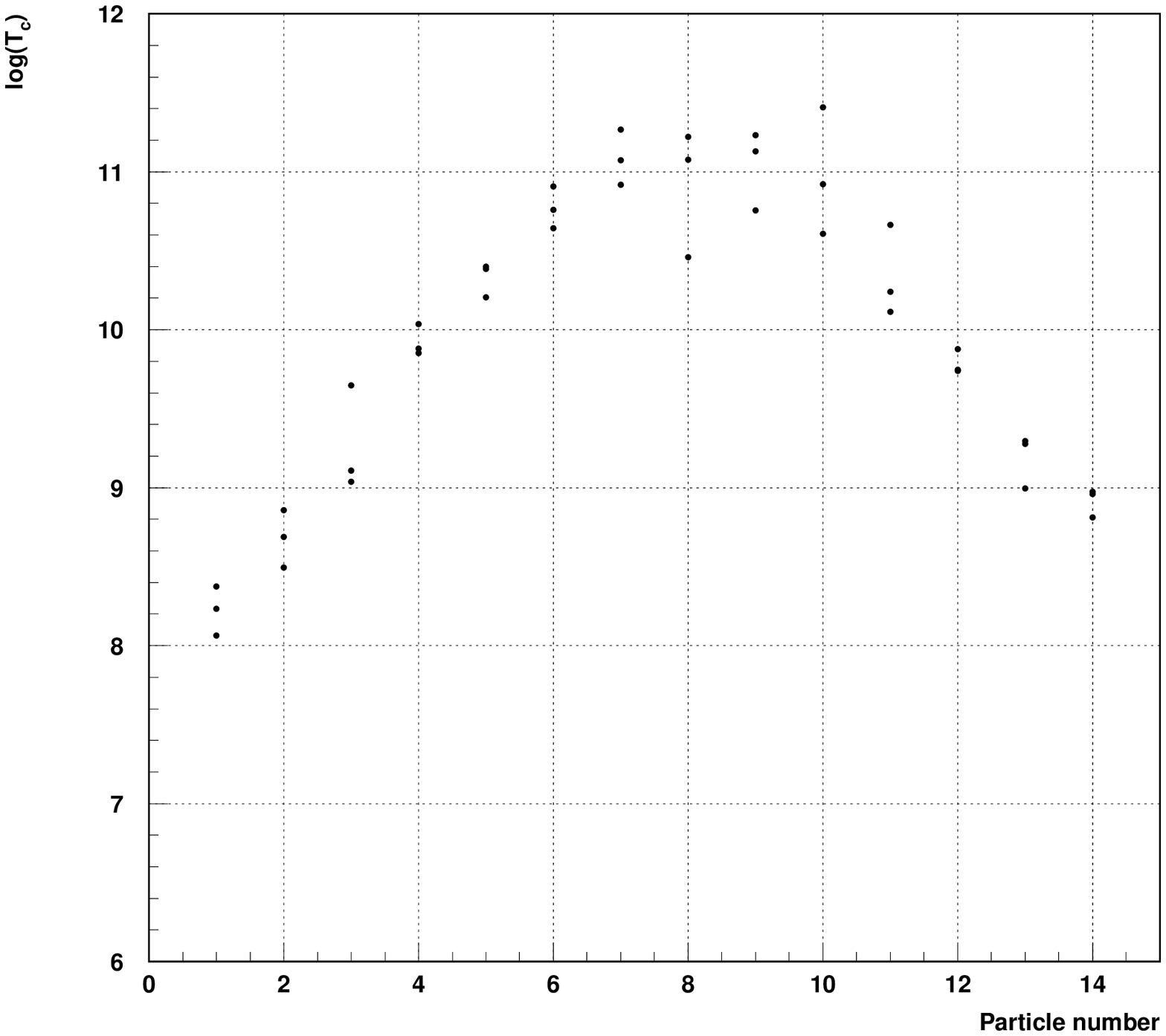}} \caption{Plot
of $\log{T_c}$ against particle number.} \label{particle.fig}
\end{figure}

Figure \ref{particle.fig} shows results from runs on an 8 vertex
lattice  with $X = 0.9$, $\theta=\pi/4$. In any given run the two
initial states are eigenstates with the same particle number,
which varies across the runs. The plot is of $T_{c}$ (defined by
$B_8$ and $\delta = 10^{-4}$) against particle number.

The plot is consistent with expected behaviour. The fixed particle
number, $m$,  sectors have dimension $(2N)!/m!(2N-m)!$ which
increases as $m$ increases to $N$ and then decreases symmetrically
as $m$ increase further to $2N$. When the Hilbert space is larger,
we expect that convergence will take longer as it takes longer for
each state to mix and acquire amplitudes for all the different
eigenstates. We expect the plot to be symmetric because, further,
there is a duality in the models between field value 0 and field
value 1.

\begin{figure}[p]
\epsfxsize=11cm \centerline{\epsfbox{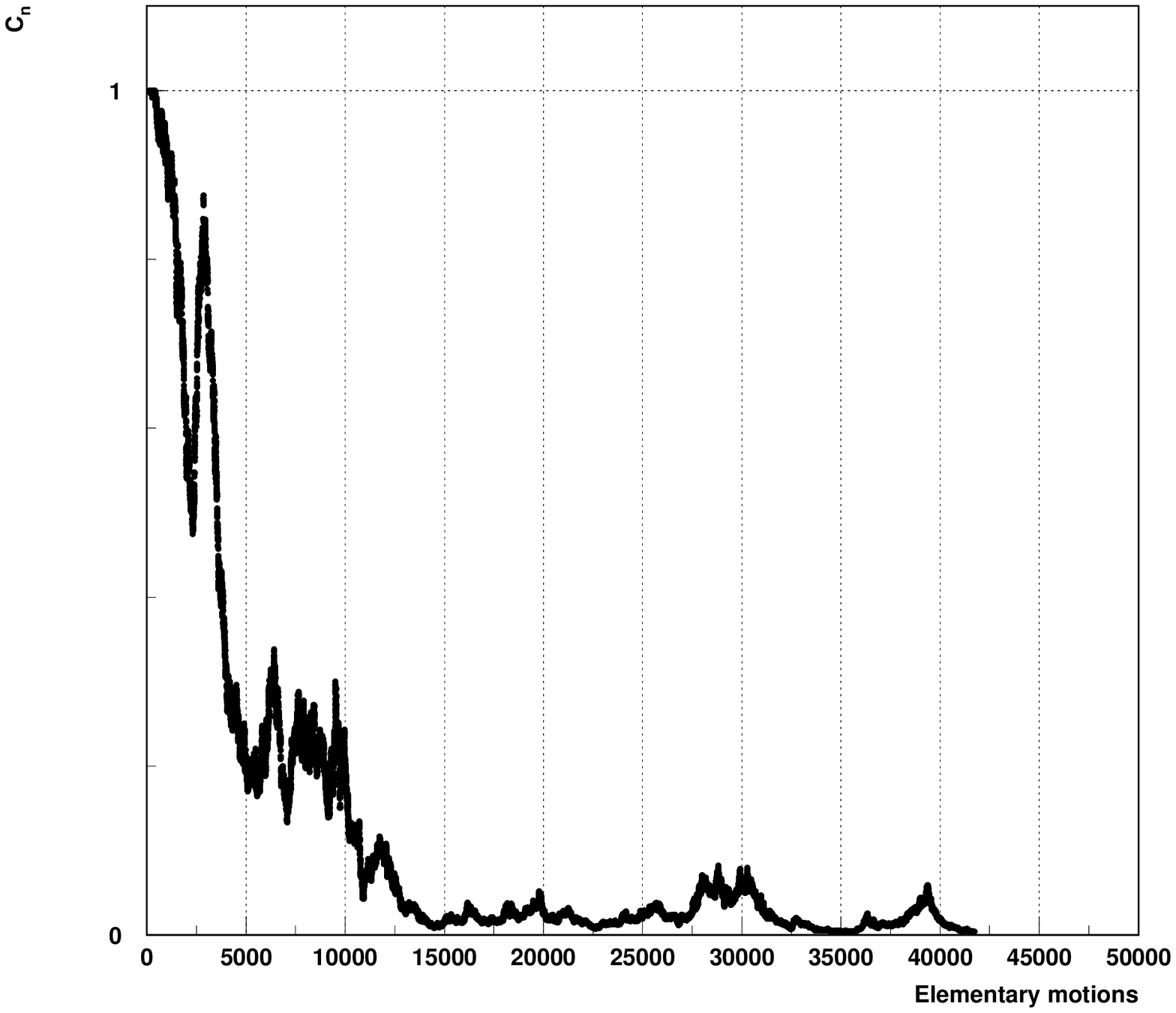}}
\caption{Plot of $C_{n}$, against the number of elementary motions, for
$X=0.95$, $\theta= 0.1\pi$ and two different one-particle
eigenstates for initial states. The data is from the same run
as shown in figure \ref{B10.fig}}
 \label{cn.fig}
\end{figure}

We checked our results by taking several runs and calculating the
quantity

\be C_n = 1 - \Vert\bra \psi^1_n, \alpha(n)| \psi^2_n, \alpha(n)
\ket \Vert^2 \ee
and comparing it to the quantity $B_{10}(t)$ for
the same run (recall that the way $B_{10}(t)$ is defined, there is
one value for every tenth lattice time coordinate, so there are 80
times as many $C_n$ data as $B_{10}$ data). Figure \ref{cn.fig} is
a $C_n$ plot of the run shown in \ref{B10.fig}. This is a more
direct check of the conjecture and in the future we would want to
redo our analysis using this method. 

\begin{figure}[p]
\begin{minipage}[b]{0.5\linewidth} 
\centering
\includegraphics[width=6cm]{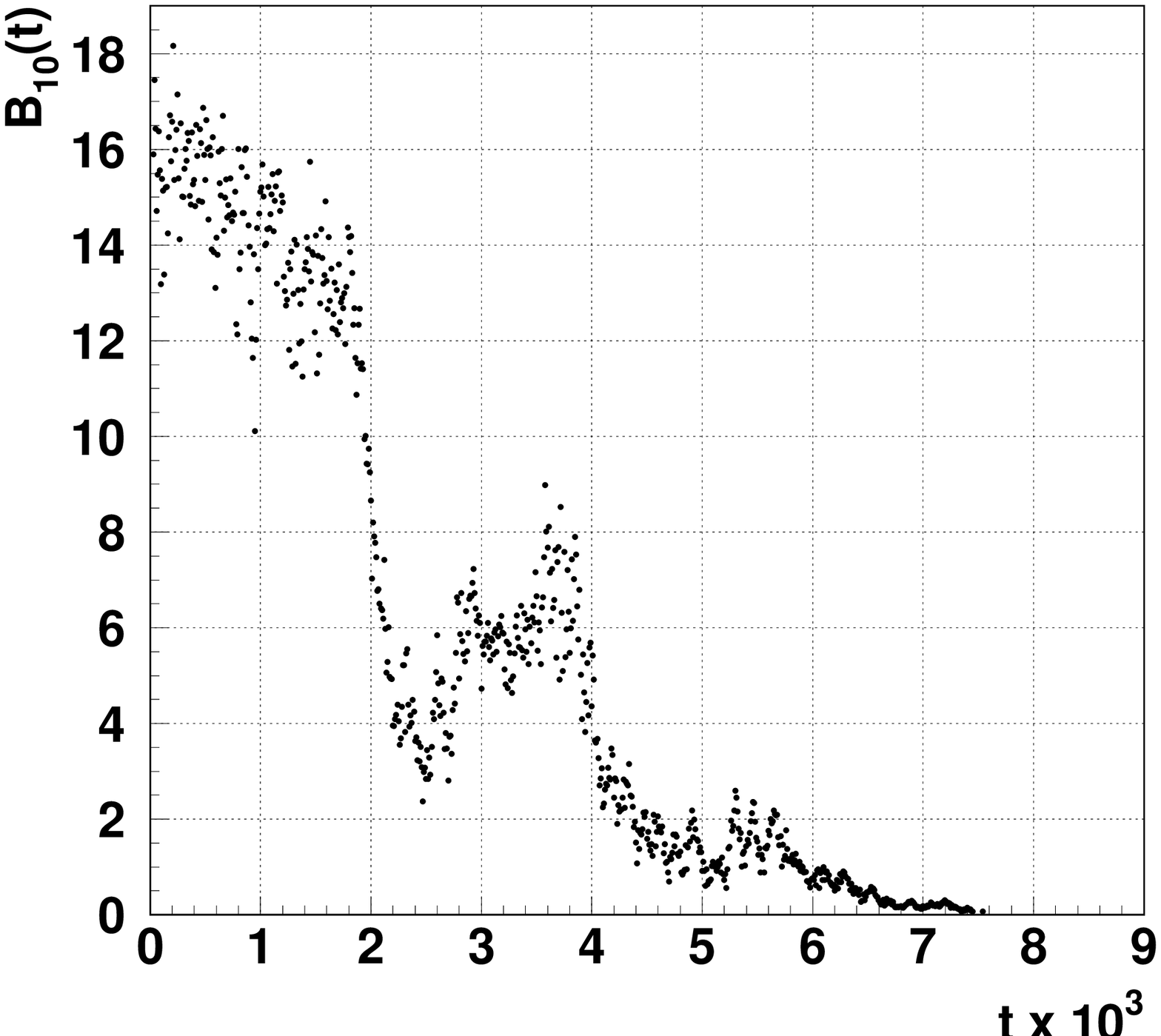}
\caption{Plot of $B_{10}(t)$, against lattice time $t$, for
$X=0.95$, $\theta= 0.1\pi$ and two different one-particle
eigenstates for initial states.}
\label{b10_2.fig}
\end{minipage}
\hspace{0.1cm} 
\begin{minipage}[b]{0.5\linewidth}
\centering
\includegraphics[width=6cm]{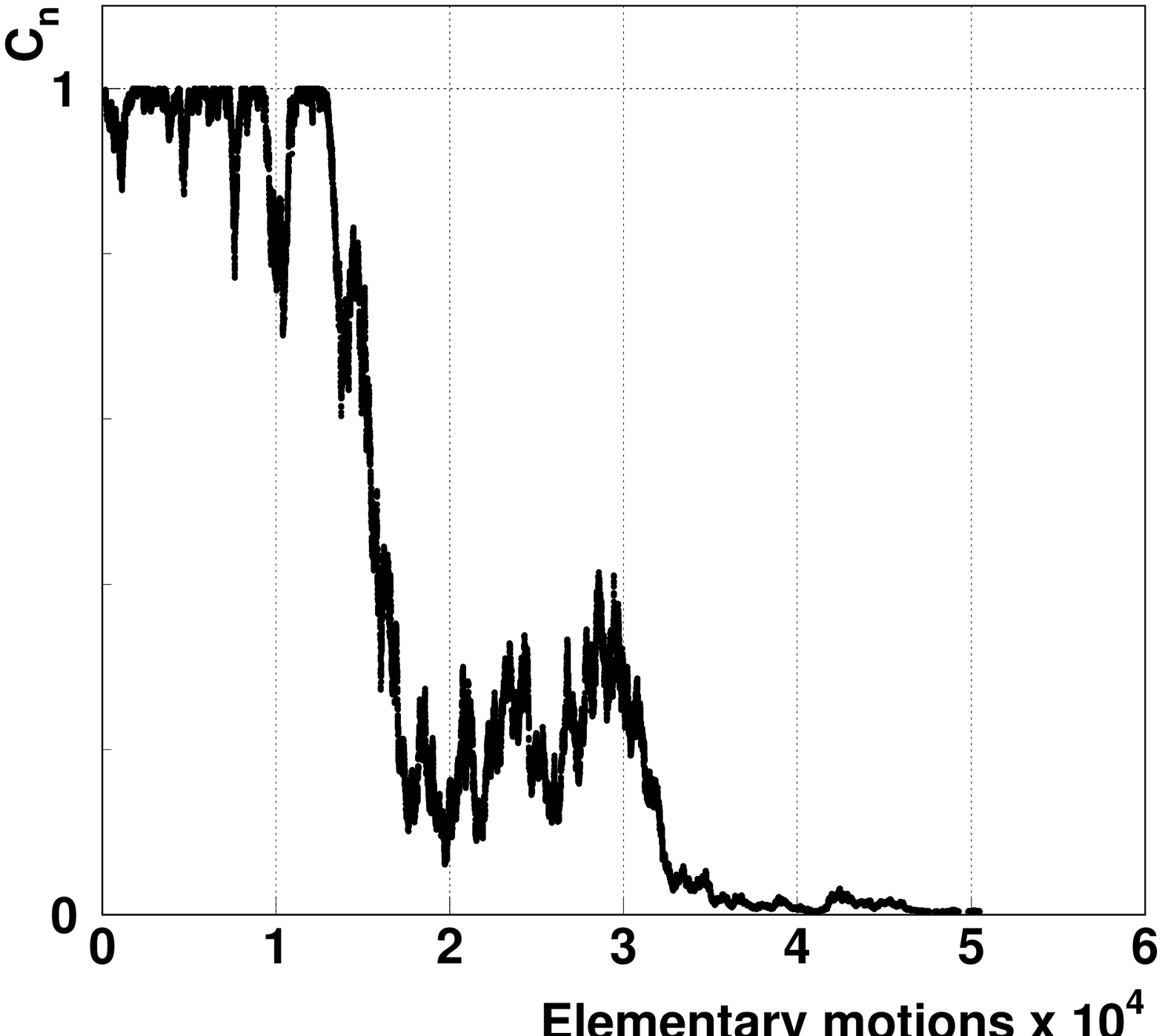}
\caption{Plot of $C_{n}$, against the number of elementary motions.
The data is from the same run as shown in
figure \ref{b10_2.fig}.
} \label{cn_2.fig}
\end{minipage}
\end{figure}

\begin{figure}[p]
\begin{minipage}[b]{0.5\linewidth} 
\centering
\includegraphics[width=6cm]{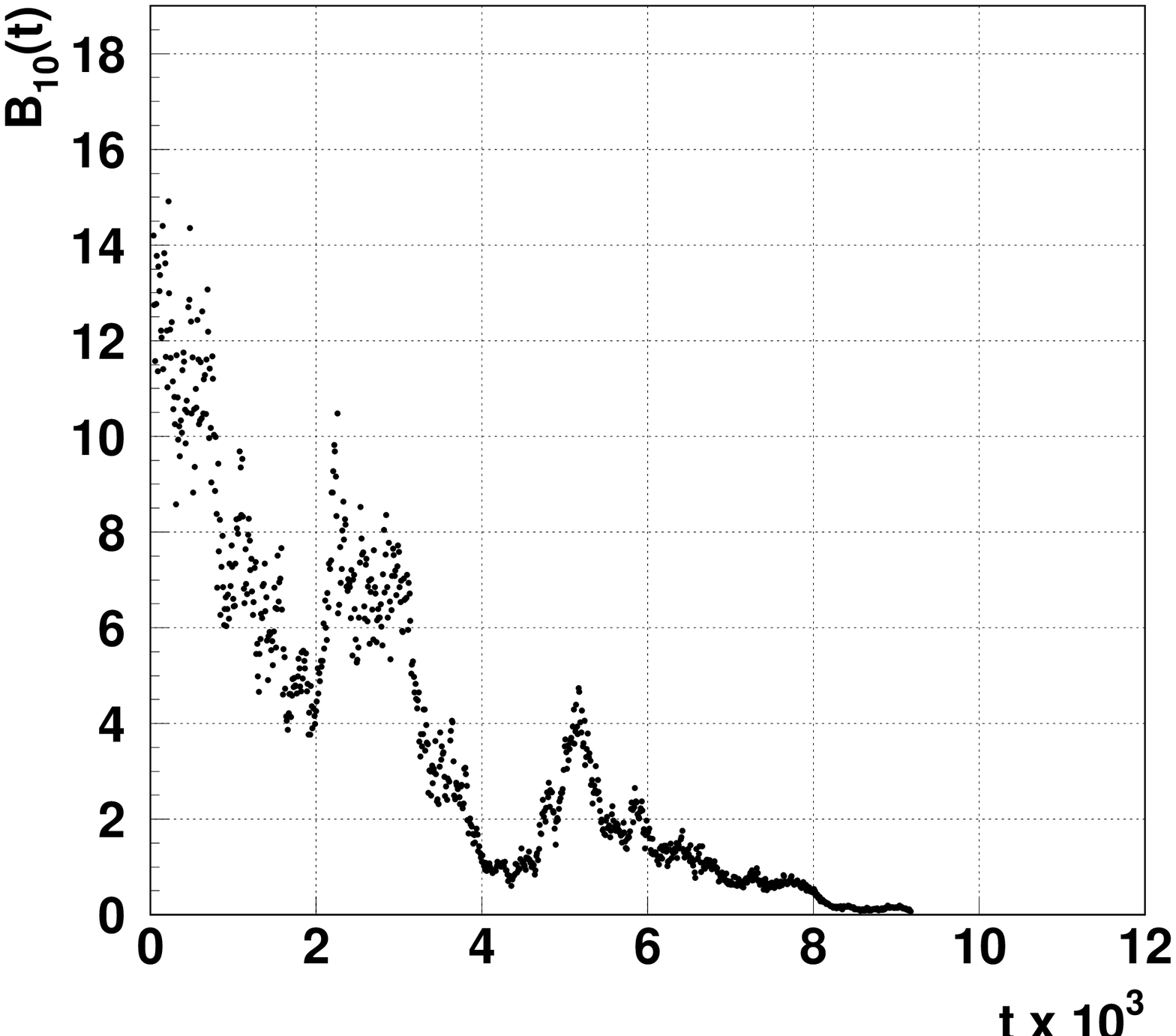}
\caption{Plot of $B_{10}(t)$, against lattice time $t$, for
$X=0.95$, $\theta= 0.25\pi$ and two different one-particle
eigenstates for initial states.}
\label{b10_3.fig}
\end{minipage}
\hspace{0.2cm} 
\begin{minipage}[b]{0.5\linewidth}
\centering
\includegraphics[width=6cm]{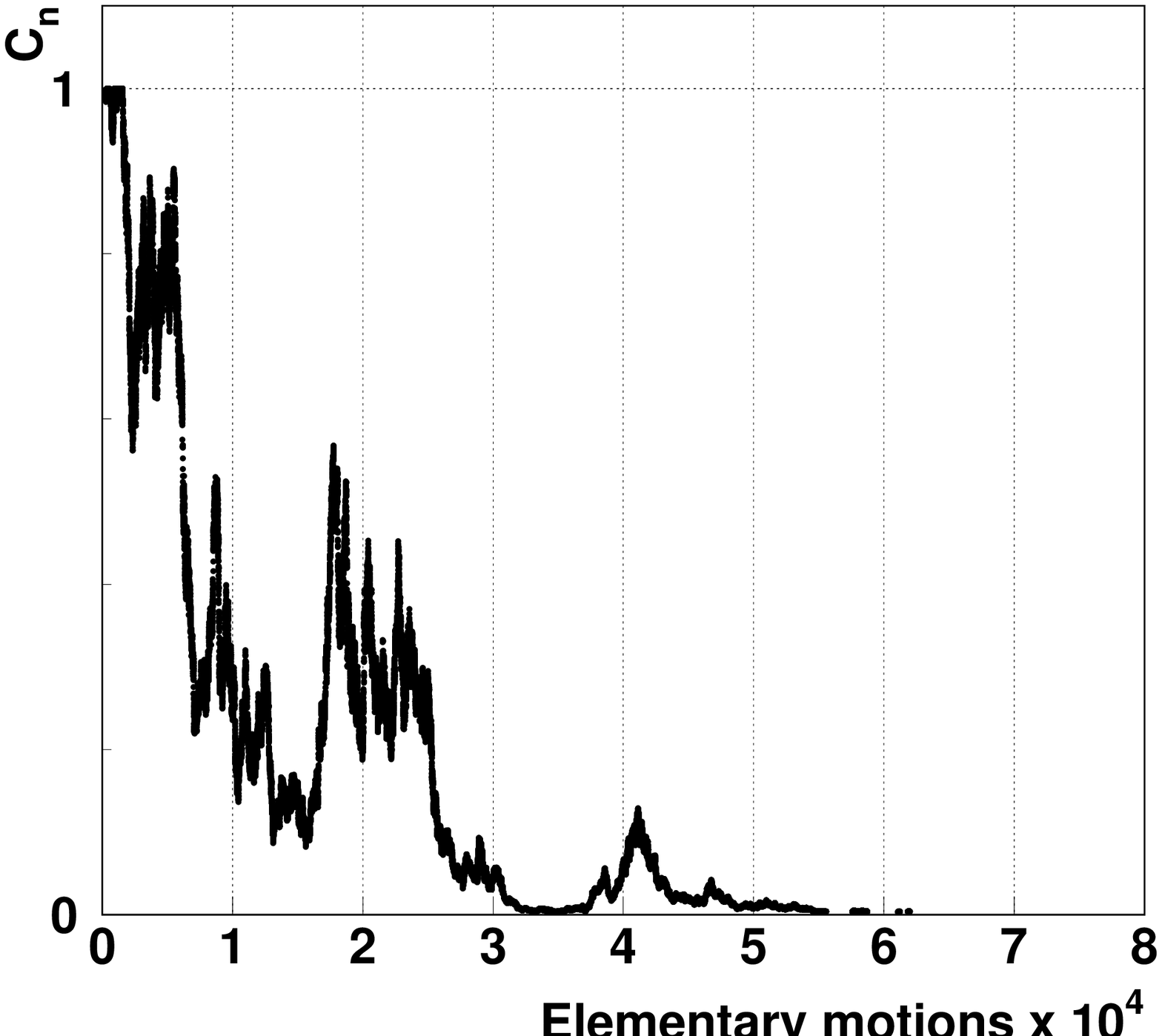}
\caption{Plot of $C_{n}$, against the number of
elementary motions.
The data is from the same run as shown in
figure \ref{b10_3.fig}.
} 
\label{cn_3.fig}
\end{minipage}
\end{figure}

However, we present more
evidence in figures \ref{cn_2.fig}, and \ref{cn_3.fig} that indicates that the results will be the same.
Indeed, even in the details of how the
convergence occurs in each run, the behaviours of the measures
$B_{10}(t)$ and $C_n$ match each other very well. On noting that the number of elementary motions is
8 times the lattice time, it can be seen that
the main features of the two types of plot
are well matched in time. 
Figures
\ref{b10_2.fig}, \ref{cn_2.fig} show a plot of  $B_{10}(t)$ and
$C_n$ data for a run with the same initial states and parameters
as for the simulation whose data is shown in figures \ref{B10.fig}
and \ref{cn.fig}, while figures \ref{b10_3.fig}, \ref{cn_3.fig}
show a plot of  $B_{10}(t)$ and $C_n$ data for a run with the same
states and parameters as for \ref{b10_2.fig} and \ref{cn_2.fig}
but a different $\theta= 0.25\pi$.

\section{DISCUSSION}

We can state the import of the conjecture we have made thus: given
some particular field history from $t = -\infty$ to $t=0$ then
there is a physical probability distribution on the field
histories for $t>0$ which we can express conveniently in the form
\eqref{eq:prob_coll}, using a quantum state at $t=0$ which is
precisely specified by the past field history. If one could 
discover an algorithm for transforming the data in the field 
history directly into the probability distribution then one
would have built a model in terms only of the field variables
which makes no reference to a quantum state. 

We have presented evidence for this conjecture. The falloff seen in figures
\ref{B10.fig} and \ref{cn.fig}-\ref{cn_3.fig} suggest the stronger
conjecture that there is a time scale $T_{c}$ such that even if we
know only the field history from $t = -T_{c}$ to $t=0$ then we can
construct a quantum state that gives the correct predictions FAPP.
We would like to study this further by investigating the
dependence of $T_c$ on the degree of convergence $\delta$.

For the purposes of this paper we chose $\delta = 10^{-4}$ to
define $T_c$ and we presented evidence that $T_{c}$ is of order
$\epsilon^{-2}$ as $\epsilon \rightarrow 0$. In a continuum limit
where the lattice spacing $a\rightarrow 0$ and $\epsilon =
O(\sqrt{a})$ then the physical convergence timescale, $aT_{c}$
would remain finite and the dynamics would be approximately
Markovian for time scales larger than this.

It would be valuable to check all our results by redoing the
simulations and calculating, instead of $B_m(t)$, $C_n$ on each
sampled surface and examining how it tends to 0 as we did for some
runs described in the last section. Improvements on our methods
would include calculating and implementing the covariant evolution
rule for surfaces which would make our sampling of surfaces
uniform. We would like to analyse quantitatively the dependence of
$T_{c}$ on the dimension of the particle number sector implied by
the results shown in figure \ref{particle.fig}.

Results with different types of R-matrices as well as general
initial states are still to be investigated. In particular,
further evidence for the conjecture  can be obtained by choosing
 pair-particle conserving matrices, as well as general matrices with
 no conservation laws.

We stress that the analysis and simulations presented in this
paper are at a rather mathematical level. The question of physics
has not been addressed. This would involve the settling of the
issue of the competition between the R-matrices and the hits in
the collapse of superpositions of eigenstates
\cite{Dowker:2004zn}. This bears on the conclusions of the current
paper. The physically interesting range of parameters is when
$\epsilon$ is very small and $\theta$ is also small so that
``microscopic'' superpositions persist for a long while but
eventually collapse. In this regime, the hits are very gentle and
the ``mixing'' of the Hilbert space by the R-matrices is slow.
Investigating this regime is essential if we are to draw
physically relevant conclusions about collapse models of this
sort.

Finally we extend our conjecture to all collapse models. It would
be interesting to study it in other cases such as the GRW model
and Di\'osi's single particle model.

\section{ACKNOWLEGMENTS}
 We thank Lajos Di\'osi, Shelly Goldstein, Joe Henson, Philip
Pearle, Ian Percival and Rafael Sorkin for helpful discussions. We
also thank Paul Dixon for help with the simulations. Most of the
work presented here was completed while FD was visiting the
Perimeter Institute (PI) for Theoretical Physics in Waterloo,
Canada. IH also spent a short visit there and both authors are
grateful for the Institute's support.
\bibliography{refs} \bibliographystyle{JHEP}
\newpage
\listoffigures
\end{document}